\documentclass[
    ,final            
  ]
  {aipproc}

\layoutstyle{8x11single}

\begin{document}

\title{Latest Flow Results from PHENIX at RHIC}



\author{E. Richardson for the PHENIX Collaboration \footnote{email: ericr@umd.edu}}{
  address={University of Maryland, College Park, MD 20742, USA},
}

\begin{abstract}
 At the Relativistic Heavy Ion Collider (RHIC), key insights into the bulk properties of the hot and dense partonic matter arise from the study of azimuthal anisotropy ($v_{2}$) of the produced particles.  These insights include indicating the matter undergoes rapid thermalization and behaves hydrodynamically at low $p_{T}$.  Recently a low energy scan ($\le \sqrt{s_{NN}}$ = 62.4 GeV) began at RHIC to search for the QGP critical point, where a change in $v_{2}$ from higher energies could play a key role in its identification.  Additionally, higher order flow harmonics have recently been shown to provide constraints on initial geometry fluctuations.  Discussed here are some of the latest low energy and higher order flow results from PHENIX.
\end{abstract}

\maketitle


\section{}

Studies of azimuthal anisotropy, or flow, in heavy-ion collisions have resulted in many key insights into the properties of the quark gluon plasma (QGP).  In particular, the $2^{nd}$ harmonic flow signal, $v_2$, has revealed that the medium undergoes rapid thermalization and behaves hydrodynamically at low transverse momentum ($p_T$).   

\begin{figure}
  \includegraphics[width=0.99\textwidth]{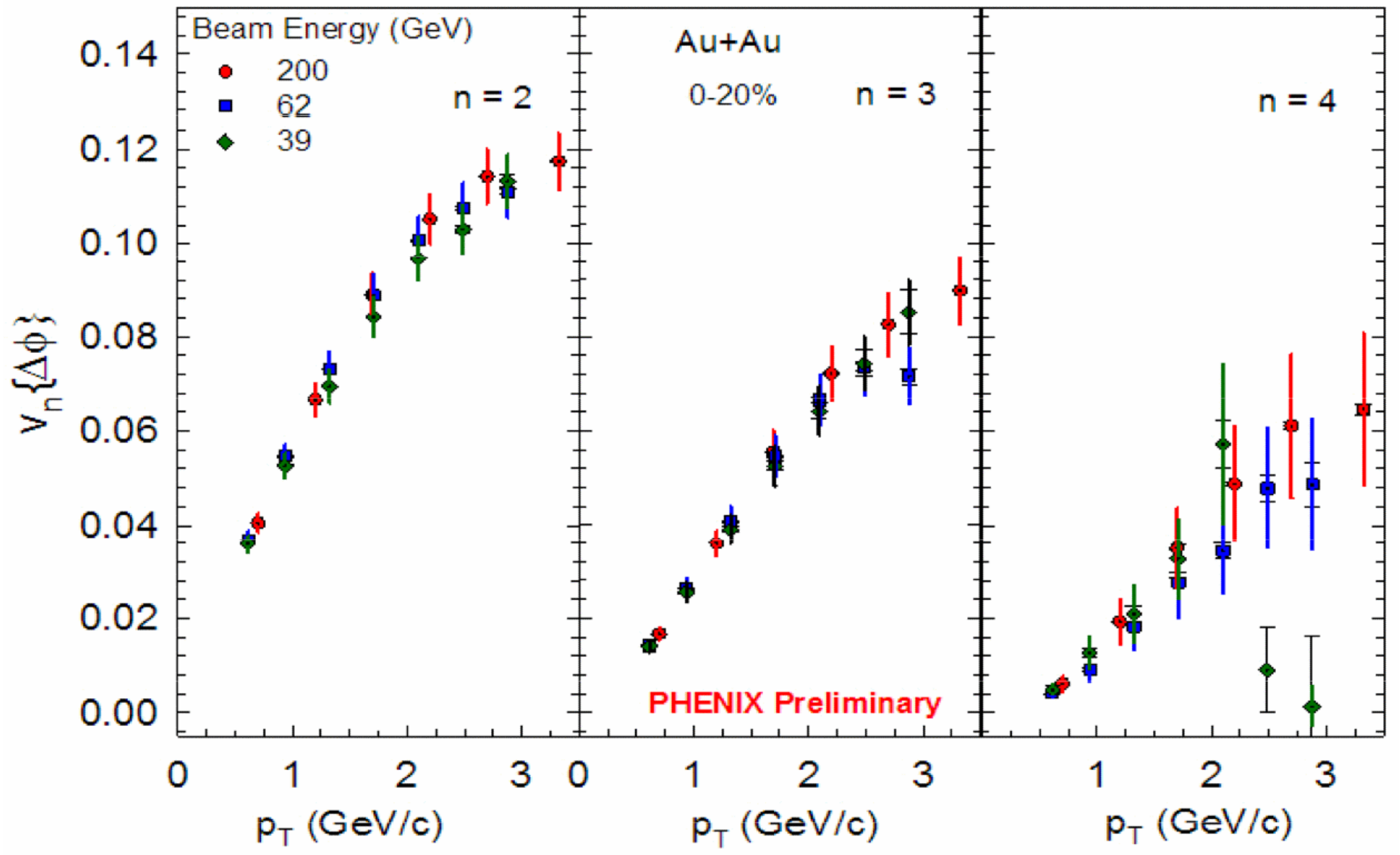}
  \caption{
  	\label{fig:vn_pT_BES}
  	PHENIX preliminary $v_n(p_T)$ for 200, 62.4 and 39 GeV beam energies using a centrality range of 0-20\%.  Despite the significant differences in beam energies the flow signals are consistent within the different harmonics, indicating similar hydrodynamic properties.}
\end{figure}

\begin{figure}
	\centering
	\includegraphics[width=0.99\textwidth]{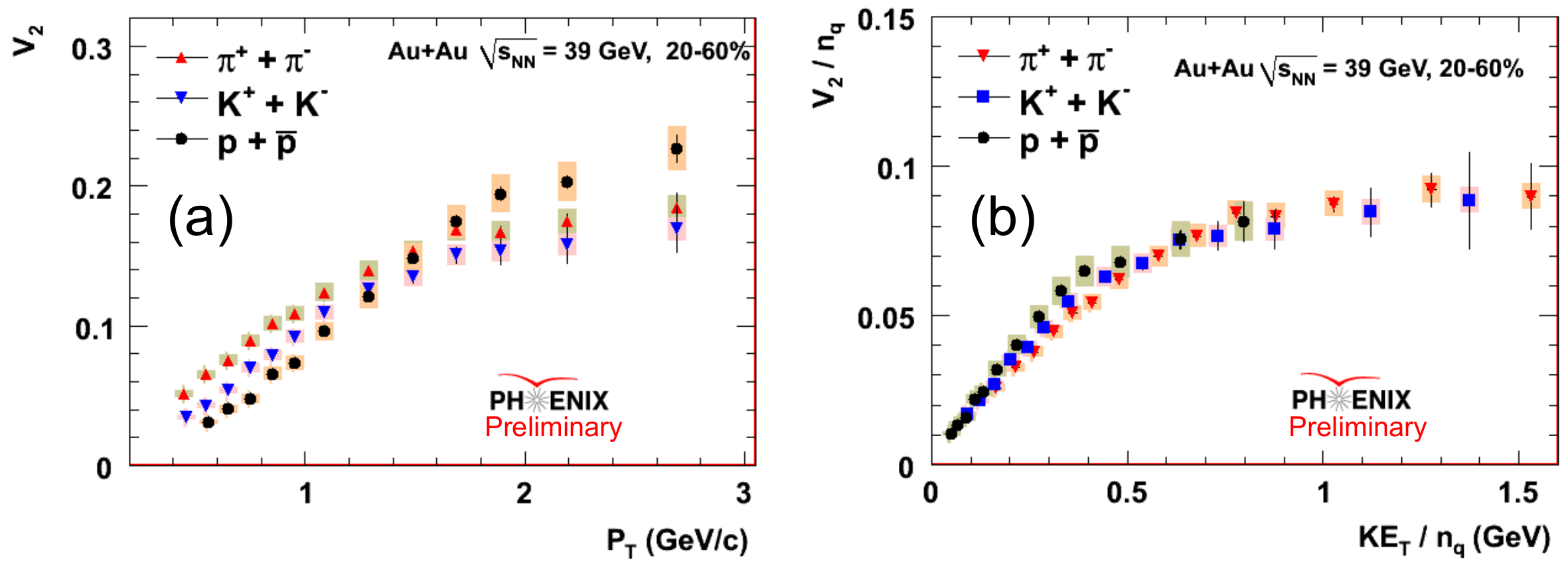}
	\caption{ 
		\label{fig:v2_PID_39GeV}
		($a$) shows PHENIX preliminary $\pi^{\pm}$, $K^\pm$ and (anti)proton $v_2(p_T)$ for 20-60\% centrality Au+Au collisions at 39 GeV, while ($b$) shows $v_2/n_q(KE_T/n_q)$ for the same identified particles.  Except for a small deviation for (anti)protons at $KE_T/n_q\approx0.4$ GeV, both ($a$) and ($b$) demonstrate similar behavior as 200 GeV collisions~\cite{bib:ppg062}, indicating similar medium properties at the two energies. }
\end{figure}

Recently RHIC began a low energy scan ($\le \sqrt{s_{NN}}$ = 62.4 GeV) to search for the QGP critical point, where a change in the flow signal could mark its location.   Figure~\ref{fig:vn_pT_BES} shows PHENIX preliminary $v_n(p_T)$ for $\sqrt{s_{NN}}$ = 200, 62.4 and 39 GeV data where, despite up to a factor of $\sim$5 difference in beam energies, the flow signals are consistent within each harmonic, indicating the hydrodynamic properties are similar within this beam energy range.  Further evidence of similar behavior is seen in Fig.~\ref{fig:v2_PID_39GeV}, where identified particle $v_2$ is shown for 39 GeV data.  In ($a$) the same mass scaling below and the same baryon/meson splitting above $p_T\approx$ 1.5 GeV/c is seen as in the 200 GeV data~\cite{bib:ppg062}.  Furthermore, ($b$) shows the same $n_q$ scaling behavior, indicating partonic level interactions, as seen with 200 GeV data, however, there is a small scaling discrepancy around $KE_T/n_q\approx0.4$ GeV.

\begin{figure}
	\centering
	\includegraphics[width=0.95\textwidth]{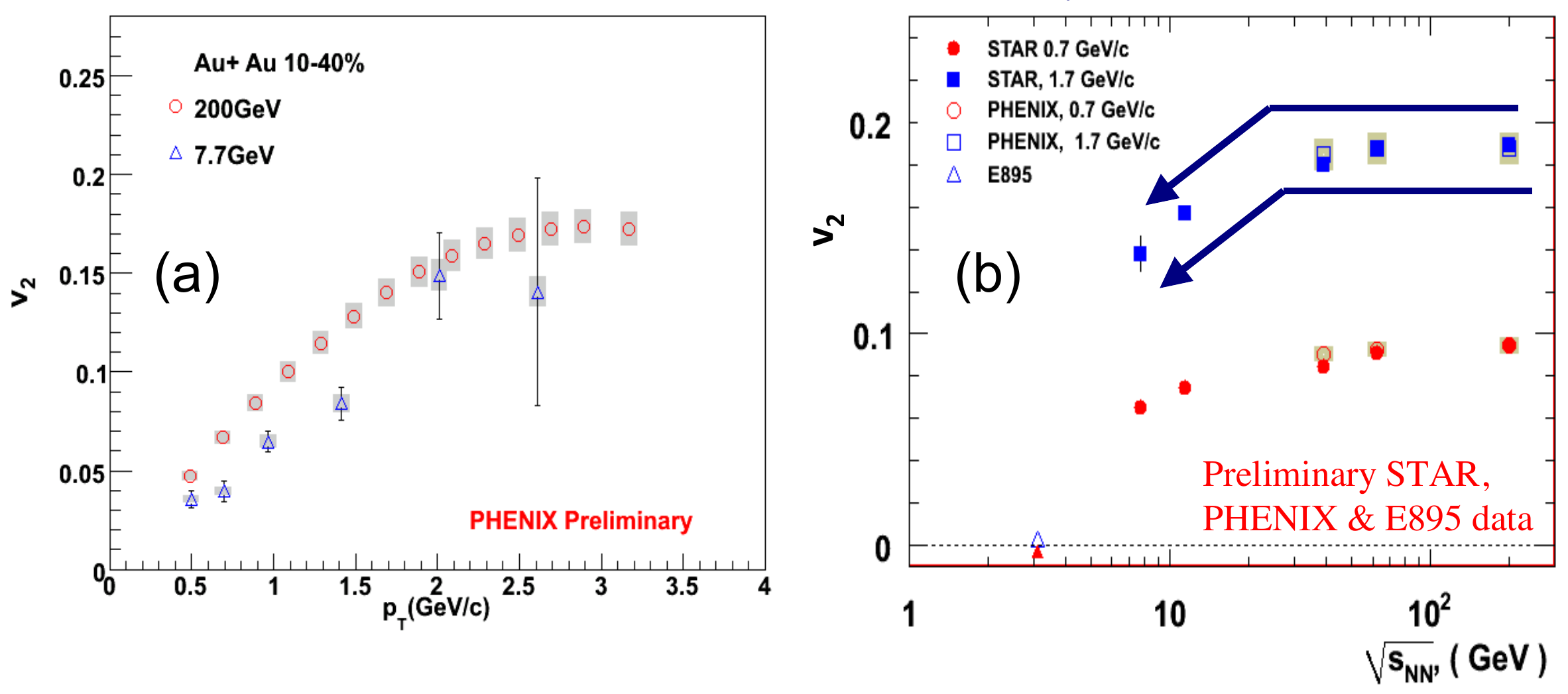}
	\caption{ 
		\label{fig:v2_pT_7GeV}
		($a$) shows the significant difference between $\sqrt{s_{NN}}$ = 7 and 200 GeV data for $v_2(p_T)$, indicating a difference in the medium's properties.  ($b$) shows $v_2(\sqrt{s_{NN}})$ at specific $p_T$ values for various beam energies from PHENIX, STAR and E895.  Here a flattening of the signal is seen above 39 GeV, indicating similar medium properties, while a decrease in signal begins in a transition region somewhere below this energy, indicating a change in the medium's properties. }
\end{figure}

Conversely, if the $v_2$ of $\sqrt{s_{NN}}$ = 7 GeV data is compared to 200 GeV data, a significant difference is seen, as shown in Fig.~\ref{fig:v2_pT_7GeV}($a$), indicating a change in the mediums properties.  Displayed another way, Fig.~\ref{fig:v2_pT_7GeV}($b$) shows $v_2(\sqrt{s_{NN}})$, where the signal flattens between 39 and 200 GeV, but starts decreasing somewhere below 39 GeV.  Does this indicate the medium is changing from partonic to hadronic interactions?  Can this help reveal the critical point?  Investigations are continuing in this transition region, including the collection of new data at 19.6 and 27 GeV during the 2011 RHIC run, which will help in answering these questions.

\begin{figure}
	\centering
	\includegraphics[width=0.95\textwidth]{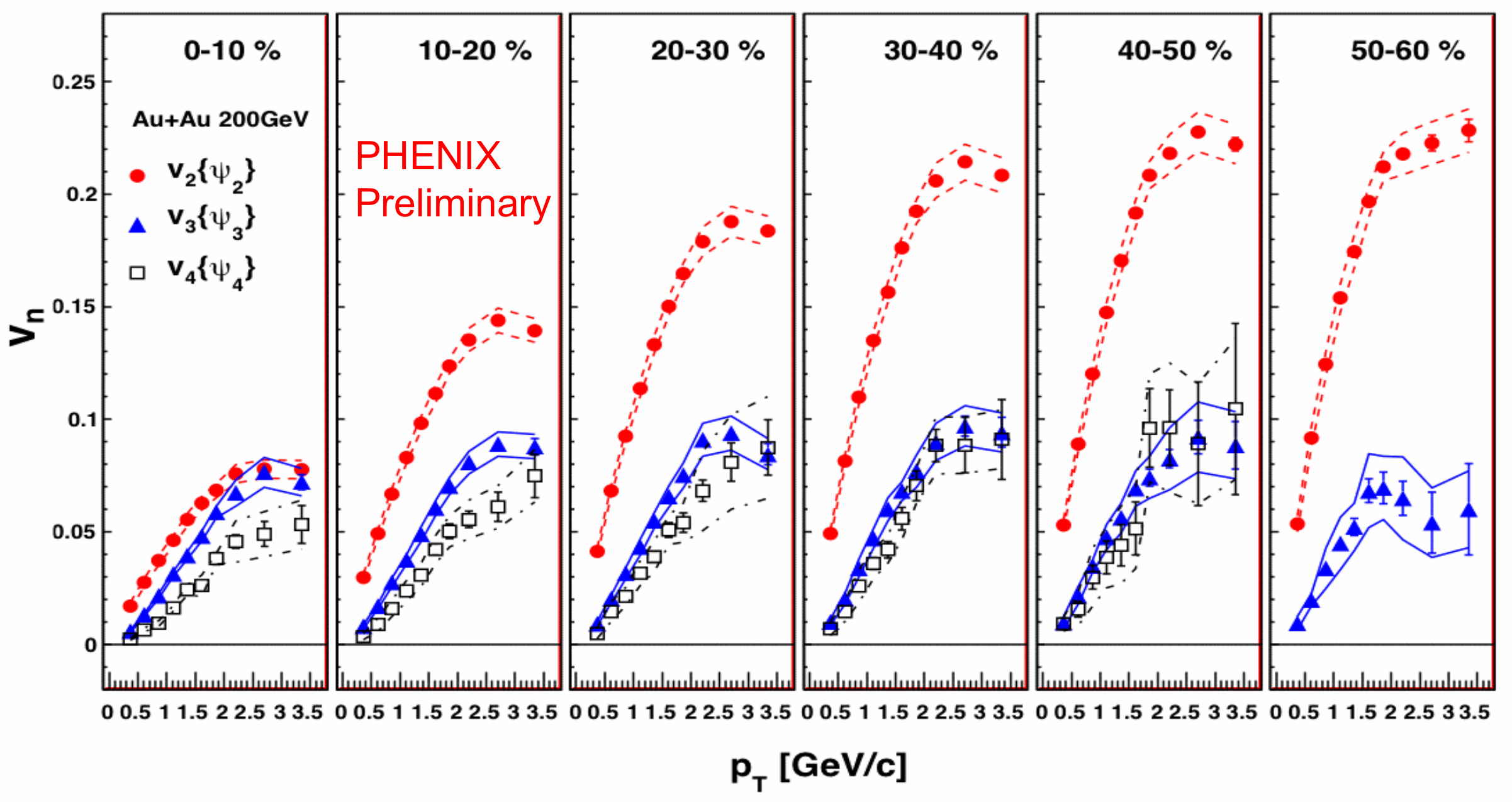}
	\caption{ 
		\label{fig:vn_pT_200GeV}
		PHENIX preliminary $v_n(p_T)$ of Au+Au 200 GeV data for different centrality ranges and within a pseudorapidity ($\eta$) region of $|\eta|\le$ 0.35.  Statistical and systematic errors are shown by the bars and dashed lines, respectively. }
\end{figure}

Recently, higher order harmonic measurements, such as $v_3$ and $v_4$, have started revealing insights about the medium's initial geometry.  Traditionally, the medium has been thought of as having a smooth texture where the odd harmonics ($v_1$, $v_3$, $v_5$...) would cancel out and be consistent with zero at mid-rapidity due to symmetry.  However, recent studies~\cite{bib:triangular_flow} have shown that this may not be the case and the medium may in-fact be ``chunky'' due to fluctuations in the participant nucleon's collective geometry, causing fluctuations in the flow signal and resulting in the odd harmonics persisting.  Figure~\ref{fig:vn_pT_200GeV} shows that indeed the $v_3$ signal is significantly positive at mid-rapidity with a weak centrality dependence, which are both consistent with a chunky initial geometry leading to fluctuations.  Additionally, $v_4$, shown here using the $4^{th}$ harmonic event plane ($\Psi_4$), is $\sim$2x larger than when using $\Psi_2$~\cite{bib:ppg098} due to $\Psi_4$ originating not only from the eccentricity and pressure gradients that drive $\Psi_2$, but also the same fluctuations as $v_4$, thus resulting in a stronger correlation and larger signal.

To summarize, in an effort to find the critical point PHENIX has measured $v_2$, $v_3$ and $v_4$ as a function of $p_T$ for $\sqrt{s_{NN}}$ = 200, 62.4 and 39 GeV, where similar signals are seen within each harmonic, indicating similar medium properties.  However, a significant decrease in $v_2(p_T)$ is observed when comparing these higher energies to $\sqrt{s_{NN}}$ = 7 GeV, indicating differing medium properties.  To further investigate, data from other beam energies within this transition region has recently been collected.  Additionally, PHENIX has measured the flow of higher order harmonics, which should provide additional constraints on initial geometry fluctuations and hydrodynamical evolution.


\bibliographystyle{unsrt}  

\bibliography{1K552_Richardson}

\begin{thebibliography}{1}

\bibitem{bib:ppg062}
A.~Adare et~al.
\newblock Scaling properties of azimuthal anisotropy in {A}u+{A}u and {C}u+{C}u
  collisions at $\sqrt{s_{NN}}=200$ {G}e{V}.
\newblock {\em Phys. Rev. Lett.}, 98:162301, 2007.

\bibitem{bib:triangular_flow}
B.~Alver et~al.
\newblock Triangular flow in hydrodynamics and transport theory.
\newblock {\em Phys. Rev. C}, 82:034913, 2010.

\bibitem{bib:ppg098}
A.~Adare et~al.
\newblock Elliptic and hexadecapole flow of charged hadrons in {A}u+{A}u
  collisions at $\sqrt{s_{NN}}=200$ {G}e{V}.
\newblock {\em Phys. Rev. Lett.}, 105:062301, 2010.

\end{thebibliography}

\end{document}